\documentclass[
reprint,
 amsmath,amssymb,
 aps,
]{revtex4-2}

\usepackage{xcolor}
\usepackage{tabularx}
\usepackage{threeparttable}
\usepackage{booktabs}

\usepackage{float}
\usepackage{mathtools}
\usepackage{graphicx}
\usepackage{dcolumn}
\usepackage{bm}
\usepackage{upgreek}
\usepackage{lipsum, babel, xcolor}
\usepackage{amsmath, amssymb}
\usepackage{comment}
\usepackage{natbib}
\usepackage{physics}
\usepackage{tabularx}
\usepackage{booktabs}
\usepackage{hyperref}
\hypersetup{colorlinks=true,citecolor=blue}

\newcommand{\dtil}{\tilde{d}}

\newcommand{\dpar}{d_{\mathrm{par}}}
\newcommand{\dzero}{d_{0}}

\newcommand{\aeff}{\alpha_{\mathrm{eff}}}
\newcommand{\awg}{\alpha_{\mathrm{wg}}}
\newcommand{\azero}{\alpha_{0}}
\newcommand{\aell}{\alpha_{\ell}}

\newcommand{\GAFC}{\Gamma_{\mathrm{AFC}}}

\newcommand{\Yb}{{}^{171}\mathrm{Yb}^{3+}}
\newcommand{\YSO}{\mathrm{Y_2SiO_5}}
\newcommand{\Topt}{T_2^{\mathrm{opt}}}

\begin{document}

\preprint{APS/123-QED}

\title{Scalable Hybrid Device Architecture on Thin-Film Lithium Tantalate for Long Distance Quantum Network Nodes with Atomic Frequency Comb Quantum Memories}

\author{Ayed Sayem\textsuperscript{1}, Towsif Taher\textsuperscript{2}}
 
\affiliation{%
\textsuperscript{1}Nokia Bell Labs, USA\\
\textsuperscript{2}Quantum Technologies, Department of Applied Physics, University of Geneva
}

\date{\today}

\begin{abstract}

Scalable photonic circuits are essential for realizing practical quantum networks beyond proof-of-principle demonstrations. Although diamond-based integrated photonics has been extensively investigated over the past two decades for quantum networking applications, its integration into scalable photonic platforms remains challenging. Quantum memories based on rare-earth-ion atomic frequency combs (AFCs) offer an attractive alternative; however, most demonstrations to date have relied on bulk free-space crystals, limiting their scalability for practical implementations. In this work, we systematically investigate the potential for hybrid integration of AFC-based quantum memories with a scalable integrated photonic platform. We propose a device architecture combining thin-film lithium tantalate (TFLT) nanophotonic circuits with ytterbium-doped yttrium orthosilicate (Yb:YSO) crystals for long-distance quantum communication. Our analysis shows that the hybrid TFLT–YSO platform offers a promising and scalable pathway toward integrated quantum memories and, ultimately, large-scale quantum networks.

\end{abstract}

\maketitle

\section{Introduction}

Quantum computing is rapidly advancing across several hardware platforms, including superconducting circuits~\cite{blais2004cavity,wallraff2004strong,houck2007single,arute2019quantum}, trapped ions~\cite{cirac1995ions,monroe1995logic,wrigh2019benchmarking}, and photonic qubits~\cite{knill2001linear,psiquantum2025manufacturable}. Rapid progress in quantum computing across superconducting, trapped-ion, photonic, and related platforms has fueled the urgency for a true quantum communication system. In particular, Shor's algorithm showed that a sufficiently large fault-tolerant quantum computer could efficiently solve integer factorization and discrete-logarithm problems, threatening widely deployed public-key cryptosystems such as RSA and elliptic-curve cryptography~\cite{shor1994,rivest1978,nistpqc2024}.\,While post-quantum cryptography provides an essential software-level response, quantum communication and quantum repeater networks offer a true hardware-level quantum mechanically protected route for distributing quantum states and cryptographic keys over long distances. A central requirement for such networks is the ability to distribute entanglement at high rate and high fidelity beyond the direct-transmission limit imposed by optical loss~\cite{briegel1998,sangouard2011}. In the United States, a major thrust in solid-state quantum-network hardware has focused on diamond color centers, particularly NV and SiV centers, because they combine optically addressable spin qubits, long-lived memory registers, and compatibility with nanophotonic cavity interfaces~\cite{ruf2021,nguyen2019diamond,knaut2024}. This direction has produced some of the most advanced demonstrations of quantum-network nodes, including nanophotonic SiV-based memory nodes interfaced with telecom fiber links~\cite{knaut2024}. These results make diamond one of the central material platforms for developing practical quantum networks. At the same time, diamond remains a difficult material platform for scalable photonic integration. Challenges include deterministic creation and placement of color centers, maintaining narrow optical linewidths near nano-structured surfaces, achieving high-yield nano-fabrication, and most importantly integrating low-loss photonic circuits with efficient fiber interfaces~\cite{rodgers2021,ruf2021}. 

Beyond vacancy-center-based memories, another attractive choice is rare earth ion based atomic-frequency-comb (AFC)-based memories in rare-earth-ion-doped solids, because they naturally support temporal, spectral, and spatial multiplexing while preserving optical coherence in a solid-state host~\cite{afzelius2009,zhou2023}. To date, many high-performance demonstrations have relied on bulk rare-earth-doped crystals, including Yb:YSO and Eu:YSO systems that have shown large multi-mode capacity, telecom-compatible correlations, and millisecond-scale photonic-qubit storage~\cite{businger2022,ortu2022}. Most of these demonstrations have been performed with bulk crystals and rely on off-chip components, severely limiting their practical impact; in this respect, diamond-based photonic circuits are far ahead but fundamentally being limited by device yield and scalability. This motivates the central question of this work: can AFC-based quantum memories be integrated into a chip-scale photonic platform with sufficient efficiency, bandwidth, and manufacturability for long-distance quantum communication? Answering this question requires identifying the appropriate material system, device architecture, and realistic performance limits for an integrated AFC-based quantum repeater node. We ask whether rare-earth-ion based AFC memories can be integrated into a scalable photonic platform, what device architecture is required for high-rate operation, and which material systems are best suited for combining low-loss photonics, efficient optical pumping, long coherence times, and telecom compatibility. Recent demonstrations in rare-earth-doped thin-film lithium niobate (TFLN)  and related integrated platforms suggest a promising route toward compact AFC devices with broadband storage, strong optical confinement, and electro-optic control~\cite{dutta2020,dutta2023,zhong2017}. Building on these advances, we evaluate the material trade-offs, device architecture, and realistic link rates that could be achieved in an integrated AFC-based quantum repeater node. 

Our analysis and results show that, from a device perspective, rare-earth-ion based quantum memories can be integrated with photonic circuits in a scalable manner especially on stable low-loss electro-optic material platfrom on thin-film lithium tantalte (TFLT). TFLT is the only platform that full-fills the major requirements for scalable integration. In particular, integrated photonic components such as high-rate entangled photon-pair sources using judicial designs can be used to take advantage of the multi-mode nature of AFC memories. Large-scale multiplexing can be implemented using MZM meshes, while hybrid waveguide–crystal memory architectures can enable integration without requiring overly complex engineering. Such platforms are also compatible with waveguide-integrated single-photon detectors, providing a path toward compact and scalable quantum-network nodes long-distance networks.

\section{Choice of material platform for photonic integrated circuit}
A central challenge in scaling AFC-based quantum memories is identifying a photonic host platform that can simultaneously provide low-loss optical routing, strong interaction with the memory medium, practical hybrid integration, electro-optic modulation for sharp pulse generation, and high optical nonlinearity for entangled photon-pair generation. In addition, the platform should be compatible with superconducting components, including microwave resonators and superconducting nanowire single-photon detectors. These requirements are particularly stringent because many solid-state quantum memories operate at visible or near-infrared wavelengths. Representative examples include diamond color centers, such as nitrogen-vacancy (NV), silicon-vacancy (SiV), and tin-vacancy (SnV) centers, as well as rare-earth-ion-doped crystals such as Pr:$\mathrm{Y_2SiO_5}$, Eu:$\mathrm{Y_2SiO_5}$, Er:$\mathrm{Y_2SiO_5}$, and Yb:$\mathrm{Y_2SiO_5}$. The relevant optical transition wavelengths for these representative memory systems are summarized in Table~\ref{tab:memory_wavelengths}. In the following sections, we discuss the key material requirements for such a hybrid integrated architecture in detail.

\begin{table*}[t]
\centering
\caption{Representative transition wavelengths for quantum memory and spin--photon interface platforms.}
\label{tab:memory_wavelengths}

\begin{tabular}{llll}
\hline
\textbf{Platform} & \textbf{Transition} & \textbf{Wavelength} & \textbf{Reference} \\
\hline
NV center in diamond & Zero-phonon line & 637 nm & \cite{Doherty2013NV} \\
SiV center in diamond & Zero-phonon line & 737 nm & \cite{Sipahigil2016SiV} \\
Pr$^{3+}$:Y$_2$SiO$_5$ & $^3H_4 \rightarrow {}^1D_2$ & 606 nm & \cite{Afzelius2010PrYSO} \\
Eu$^{3+}$:Y$_2$SiO$_5$ & $^7F_0 \rightarrow {}^5D_0$ & 580 nm & \cite{Zhong2015EuYSO} \\
Yb$^{3+}$:Y$_2$SiO$_5$ & $^2F_{7/2} \rightarrow {}^2F_{5/2}$ & 979 nm & \cite{LagoRivera2022YbYSO} \\
Er$^{3+}$:Y$_2$SiO$_5$ & $^4I_{15/2} \rightarrow {}^4I_{13/2}$ & 1536 nm & \cite{Saglamyurek2015ErFiber} \\
Er$^{3+}$:LiNbO$_3$ & $^4I_{15/2} \rightarrow {}^4I_{13/2}$ & 1532 nm & \cite{Saglamyurek2011TmLN} \\
Tm$^{3+}$:YAG & $^3H_6 \rightarrow {}^3H_4$ & 793 nm & \cite{deRiedmatten2008TmYAG} \\
\hline
\end{tabular}
\label{tab1}
\end{table*}

\newcommand{\yescell}[1]{\textcolor{green!50!black}{#1}}
\newcommand{\nocell}[1]{\textcolor{red!70!black}{#1}}
\newcommand{\notknown}{Not known}

\begin{table*}[t]
\centering
\caption{Comparison of material platforms for integrated quantum photonic circuits.}
\label{tab:material_platform_comparison}

\begin{threeparttable}

\begin{tabularx}{\textwidth}{l c c c c c c c c}
\hline
\shortstack{\textbf{Material}\\\textbf{platform}} &
\shortstack{\textbf{Transparency}\\\textbf{window}} &
\shortstack{\textbf{Low-loss}} &
\shortstack{\textbf{Pockels}\\\textbf{coefficient}\\\textbf{pm/V}} &
\shortstack{\textbf{Nonlinearity}\\\textbf{$d$ coefficient}\\\textbf{pm/V}} &
\shortstack{\textbf{Domain}\\\textbf{engineering}}\tnote{a} &
\shortstack{\textbf{Cryogenic}\\\textbf{compatibility}}\tnote{b} &
\shortstack{\textbf{DC-bias}\\\textbf{stability}} &
\shortstack{\textbf{Stable pulse}\\\textbf{generation}} \\
\hline

Silicon &
$1.1{-}8~\mu$m \cite{Subramanian2015SiSiN} &
\nocell{No} &
\nocell{No} &
\nocell{No} &
\nocell{No} &
\nocell{No} &
N/A &
N/A \\

SiN &
$0.4{-}2.35~\mu$m \cite{Geng2024NonlinearPhotonics} &
\yescell{Yes} &
\nocell{No} &
\nocell{No} &
\nocell{No} &
\nocell{No} &
N/A &
N/A \\

SiC &
$0.4{-}5~\mu$m \cite{Powell2022SiC} &
\yescell{Yes} &
\yescell{Yes, $r\approx 0.2{-}1.5$} &
\yescell{$d\sim 10{-}20$} &
\nocell{No} &
\yescell{Yes} &
\yescell{Yes} &
\yescell{Yes} \\

TFLN &
$0.35{-}5~\mu$m \cite{Zhu2021TFLN} &
\yescell{Yes} &
\yescell{Yes, $r_{33}\approx 31$} &
\yescell{$d_{33}\approx 27$} &
\yescell{Yes} &
\yescell{Yes} &
\nocell{No} &
\nocell{No} \\

TFLT &
$0.28{-}5.5~\mu$m \cite{Yan2020LT} &
\yescell{Yes} &
\yescell{Yes, $r_{33}\approx 30$} &
\yescell{$d_{33}\approx 15$} &
\yescell{Yes} &
\yescell{Yes} &
\yescell{Yes} &
\yescell{Yes} \\

GaP &
$0.55{-}11~\mu$m \cite{Wilson2020GaP} &
Emerging &
\yescell{Yes, $r_{41}\approx 1.1$} &
\yescell{$d_{14}\sim 50{-}70$} &
\nocell{No} &
\yescell{Yes} &
\notknown &
\yescell{Yes} \\

AlN &
$0.21{-}>8~\mu$m \cite{Li2021AlN} &
\yescell{Yes} &
\yescell{Yes, $r_{33}\approx 1.0$} &
\yescell{$d_{33}\sim 4{-}6$} &
\nocell{No} &
\yescell{Yes} &
\yescell{Yes} &
\yescell{Yes} \\

AlGaAs &
$0.9{-}17~\mu$m \cite{Mobini2022AlGaAs} &
Emerging &
\yescell{Yes, $r_{41}\approx 1{-}1.5$} &
\yescell{$d_{14}\sim 100$} &
\nocell{No} &
\yescell{Yes} &
\notknown &
\yescell{Yes} \\

InGaP &
$>0.65~\mu$m \cite{Haque2024InGaP} &
Emerging &
\yescell{Yes, $r_{41}\sim 1{-}2$} &
\yescell{$d_{14}\sim 70$} &
\nocell{No} &
\yescell{Yes} &
\notknown &
\yescell{Yes} \\

\hline
\end{tabularx}

\begin{tablenotes}
\footnotesize
\item[a] Here, domain engineering refers specifically to ferroelectric domain reversal for periodic poling.
\item[b] By cryogenic compatibility, we mean cryogenic modulation compatibility.
\end{tablenotes}

\end{threeparttable}
\end{table*}
\textbf{i) Low-loss and transparency in relative optical window} 

The photonic platform must provide broadband transparency and low propagation loss across the relevant quantum-memory transitions. Silicon photonics and InP-based photonic integration have been workhorse platforms for classical high-speed, high-bandwidth communication \cite{doerr2015silicon,shi2022silicon,smit2019inp}. However, their material absorption strongly constrains operation at many visible and short-wavelength near-infrared quantum-memory transitions~\cite{poon2024visible_silicon,vurgaftman2001iii_v}.
In contrast, low-loss dielectric and ferroelectric platforms such as aluminum oxide ($\mathrm{Al_2O_3}$), silicon nitride $\mathrm{SiN}$, thin-film lithium niobate (TFLN), and thin-film lithium tantalate (TFLT) offer broad transparency from the visible to the near-infrared, making them attractive candidates for quantum-memory integration~\cite{munoz2017sin,desiatov2019visibleln,neutens2025alumina}. Other nonlinear semiconductor platforms, including GaP and InGaP, are also promising for quantum photonics because of their high refractive index and strong optical nonlinearities, but are generally less mature as ultra-low-loss passive platforms over the full visible/near-infrared range needed for heterogeneous quantum-memory systems~\cite{Wilson2020GaP,akin2024ingap}.

\textbf{ii) Integration complexity}

Beyond optical transparency, the second essential requirement is controllable hybrid integration with the quantum-memory material. This requirement is particularly challenging for diamond-based memories. Diamond has a high refractive index, $\mathrm{n \approx 2.4}$, which is larger than that of many low-loss dielectric photonic platforms. As a result, efficient coupling between diamond color centers and external photonic circuits often requires carefully engineered nanocavities, adiabatic waveguide interfaces, or hybrid transfer techniques. Recent demonstrations, including $\mathrm{GaP}$ or other III--V cavities integrated with diamond and diamond/$\mathrm{TFLN}$ heterogeneous integration, have enabled important progress toward efficient emitter--photon interfaces and multifunctional quantum photonic platforms~\cite{huang2021hybrid,chakravarthi2023hybrid,riedel2023efficient,ding2026heterogeneously,xu2026thin,abulnaga2025design}. Nevertheless, these approaches still rely on precise nanofabrication, transfer alignment, and material-specific interface engineering; therefore, wafer-scale, high-yield, and low-loss integration of diamond quantum emitters remains challenging~\cite{ruf2021diamond,rodgers2021diamond,knaut2024diamond,chakravarthi2023hybrid,riedel2023efficient}. For example, the latest demonstration of diamond integration on $\mathrm{TFLN}$ showed nearly $\mathrm{3\,dB}$ transition loss just to couple light into the $\mathrm{LN}$ waveguide. Furthermore, after the full hybrid processing flow, the propagation loss of $\mathrm{LN}$ increased to approximately $\mathrm{3\,dB/cm}$, highlighting the undeniable complexity of hybrid integration involving materials with a higher refractive index than $\mathrm{LN}$~\cite{ding2026heterogeneously}.

Rare-earth-ions doped in rare earth crystals such as $\mathrm{YSO}$, offer a different integration opportunity. $\mathrm{YSO}$ has a relatively low optical refractive index, approximately $\mathrm{n \sim 1.7\text{--}1.8}$, \cite{weber2002handbook}. This index is lower than the host PIC platforms such as $\mathrm{Si_3N_4}$, $\mathrm{TFLN}$, or $\mathrm{TFLT}$.  This significantly relaxes the complexity of hybrid integration as the optical mode can remain primarily guided in a higher-index nanophotonic waveguide while an engineered fraction of the optical mode overlaps with the rare-earth-doped $\mathrm{YSO}$ crystal. In this geometry, the light--matter interaction strength can be engineered by controlling the waveguide dimensions, the separation from the $\mathrm{YSO}$, and the interaction length. Therefore, rather than requiring the memory material itself to be fully nanofabricated into high-confinement waveguides, a hybrid $\mathrm{AFC}$ architecture could use a low-loss photonic circuit as the optical routing layer and a rare-earth-doped $\mathrm{YSO}$ crystal or thin film as the quantum-memory layer~\cite{miyazono2017er,yso2024thinfilm,zhou2023integratedmemory,yang2021photonic}.

\textbf{iii) Fast switching circuits}

One of the key requirements for a quantum photonic circuit is fast electro-optic control, which is needed for pump-pulse generation, photon routing, Bell-state measurements (BSMs), multiplexing circuits, and related operations. Unfortunately, only a few material platforms exhibit a useful Pockels effect, including LN, LT, BTO, $\mathrm{AlN}$, and $\mathrm{SiC}$. Although BTO offers one of the largest Pockels coefficients, its propagation loss is typically higher than that of other relevant integrated photonic platforms. LN and LT provide strong electro-optic responses and have enabled high-performance thin-film electro-optic modulators~\cite{wang2018integratedTFLNModulator,wang2024lithiumTantalatePICs}. In contrast, $\mathrm{AlN}$ and $\mathrm{SiC}$ have much weaker Pockels coefficients, typically on the order of $\mathrm{1{-}1.5\,pm/V}$~\cite{xiong2012lowLossAlN,liu2023aluminumAlNReview,powell2022integratedSiCModulator,wang2023pockels4HSiC}. The lack of a strong Pockels effect eliminates many other material platforms, such as $\mathrm{SiN}$ and $\mathrm{Al_2O_3}$, for applications requiring fast electro-optic switching. Importantly, the Pockels effect persists at cryogenic temperatures, so fast electro-optic control can, in principle, be maintained without adding substantial complexity when devices are operated under cryogenic conditions.

\textbf{v) Stability and high extinction pulse generation}
Device stability is a key figure of merit for a truly scalable quantum communication system. First, many critical components, such as entangled photon-pair sources, need to use resonant structures, which can drift upon illumination due to different effects such as thermo-optic (TO) and photorefractive (PR) effects. We have recently shown that TFLT is a superior material platform in terms of device stability, especially because it performs significantly better than TFLN~\cite{sayem2026high}. Another critical stability factor is the DC-bias stability of the switches and interferometer circuits needed to create the pump pulses and perform corresponding measurements, such as BSM and photon routing. Among the candidate electro-optic materials, TFLT is the only one that offers DC-bias stability at relevant wavelengths. For example, we have recently demonstrated high-bandwidth modulators near $\mathrm{1~\mu m}$ with excellent DC-bias stability \cite{sayem2026high1}. TFLN, on the other hand, drastically suffers from DC-bias instability, which in turn requires thermo-optic (TO) circuits for bias control. Although this may work for classical applications, it is a serious drawback for quantum applications, as many of these circuits need to be operated at cryogenic temperatures. High-extinction pulse generation is another key figure of merit, as such pulses are necessary in numerous cases, for example, hole burning for AFC comb generation, Bell-state measurement, pump pulses for photon-pair generation, and switching circuits for multiplexing in detection and heralding circuits. Although this is a fundamental requirement, in reality, it is very difficult to achieve in a PIC platform. TO circuits are slow and cryogenically incompatible, so they are not an option. TFLN fails to generate sharp pulses due to the PR effect \cite{sayem2026unveiling}. We have recently shown that TFLT modulators can generate sharp pulses for a wide range of pulse widths required for quantum applications, thanks to the high activation energy required for charge excitation \cite{sayem2026unveiling}.

vi) \textbf{Cryogenic compatibility and superconducting circuit integration.}
Most quantum memories require cryogenic operation. For example, $\mathrm{SiV}$ centers in diamond require millikelvin temperatures, whereas rare-earth-ion $\mathrm{AFC}$ memories in $\mathrm{YSO}$ are commonly operated at cryogenic temperatures ranging from a few kelvin down to the sub-kelvin regime~\cite{sukachev2017sivMemory,chiossi2026ultraslowSpinRelaxationYbYSO}. $\mathrm{TFLN}$ has already shown compatibility with cryogenic and superconducting quantum hardware, including waveguide-integrated superconducting nanowire single-photon detectors and superconducting microwave resonators for bidirectional microwave--optical conversion~\cite{sayem2020lithium,xu2021bidirectional}. However, an important limitation of $\mathrm{TFLN}$ is its device stability under cryogenic operation; as described earlier, and similar to room-temperature operation, $\mathrm{TFLN}$ can suffer from low-frequency bias drift, which is particularly problematic for stable electro-optic operation in quantum systems. Recent work on quantum transduction has shown that $\mathrm{TFLT}$ can provide improved bias stability compared with $\mathrm{TFLN}$ while maintaining strong electro-optic functionality, enabling stable bidirectional microwave--optical conversion over extended operation times~\cite{axline2026stable}. These results suggest that $\mathrm{TFLT}$ may offer an attractive route toward stable cryogenic photonic interfaces for quantum-memory systems. Nevertheless, further experiments are required to fully verify the performance of $\mathrm{TFLT}$ across different quantum applications, including the quantum-memory architecture proposed in this article.

In Table~\ref{tab:material_platform_comparison}, we summarize the key characteristics of different PIC material platforms relevant to quantum photonic applications. The discussion in this section provides a clear idea of what is required for the right choice of material. Silicon is ruled out because of its lack of transparency in the desired wavelength bands. SiN is ruled out because of the lack of the Pockels effect. Among the electro-optic materials, AlN, SiC, GaP, and InGaP are ruled out because of their weak electro-optic (EO) effect, along with other factors. Then we are left with TFLN and TFLT. Both TFLN and TFLT offer excellent EO properties, low loss, and strong optical nonlinearity but TFLN offers higher optical nonlinearity than TFLT. Therefore, in terms of pure performance, TFLN would be the material of choice. Unfortunately, TFLN fails because of its lack of stability, DC-bias drift, and distorted pulses, which are key limitations for a scalable system. Hence, we choose TFLT as the host PIC material for the hybrid AFC-based quantum communication link. In the rest of this paper, we present the device components on the TFLT platform and calculate the link characteristics based on the performance achievable with TFLT PICs.

\section{Choice of quantum memory platform}
A central challenge for rare-earth ion-based quantum memories is to combine the long coherence of non-Kramers ions such as $\mathrm{Eu^{3+}}$ and $\mathrm{Pr^{3+}}$ with the much larger bandwidths available in paramagnetic rare-earth systems. The $\Yb$ isotope combines an effective electronic spin $S=1/2$ with nuclear spin $I=1/2$, giving four hybridized electron--nuclear hyperfine states with GHz-scale splittings at crystallographic site~II \cite{tiranov2018spectroscopic,welinski2016high}, making $\Yb{:}\YSO$ uniquely suitable in this respect. Furthermore, its strongly anisotropic hyperfine interaction also produces simultaneous optical and microwave zero-field-first-order-Zeeman (ZEFOZ) transitions at zero field, strongly suppressing magnetic-field sensitivity \cite{ortu2018simultaneous} and enabling simultaneous broadband operation and long coherence time.

For the $^2F_{7/2}(0)\!\leftrightarrow\!{}^2F_{5/2}(0)$ transition near $979$~nm, homogeneous linewidths as narrow as $320$~Hz, corresponding to $\Topt=1/\pi\Gamma_h\approx1$~ms, have been measured in low-doped crystals \cite{chiossi2024optical}. The ground-state clock transition reaches a spin coherence time of $10.0(4)$~ms at $3.4$~K \cite{nicolas2023coherent}, while spin-state lifetimes can extend to seconds at low concentration and remain in the millisecond regime even at $10$~ppm \cite{chiossi2026optical}. Importantly, increasing the Yb concentration from $2$ to $10$~ppm does not introduce a prohibitive optical- or spin-coherence penalty: the relevant linewidths and spin lifetimes remain compatible with AFC operation, while the increased absorption density substantially relaxes the optical-depth requirements for efficient storage \cite{chiossi2024optical,chiossi2026optical}. These properties have already enabled spin-wave AFC storage for $1.2$~ms \cite{businger2020optical}, preservation of non-classical correlations across $1250$ temporal modes over a $100$~MHz bandwidth \cite{businger2022non}, and broadband AFC storage over $250$~MHz, approaching the $288$~MHz theoretical spin-wave bandwidth of the proposed single-class preparation scheme \cite{sanchez2026broadband}. More recently, entanglement distribution with storage across more than $8000$ temporal modes was demonstrated over a deployed metropolitan fiber link \cite{rodriguez2026entanglement}.

Despite these excellent metrics, further improvement in memory performance is possible and required for potential use in a deployed quantum repeater network. The measured AFC coherence lifetime of $307$--$348~\mu$s remains about three times shorter than the underlying optical coherence time of $\sim1$~ms, leaving substantial room to extend storage duration \cite{rodriguez2026entanglement,chiossi2024optical}. Moreover, the $\sim19\%$ forward readout efficiency is primarily limited by the available optical depth \cite{rodriguez2026entanglement} and capped at $\approx54\%$. This can be improved through higher absorption density and ultimately an impedance-matched cavity \cite{moiseev2010efficient}. Since increasing the $\Yb$ concentration does not prohibitively affect useful optical linewidths and spin lifetimes \cite{chiossi2026optical}, $\Yb{:}\YSO$ is a prime candidate for a hybrid quantum memory device architecture, which we explore in detail in the later sections.

\section{Device architecture}
\begin{figure*} [!ht]
    \centering
    \includegraphics[width=0.8\textwidth]{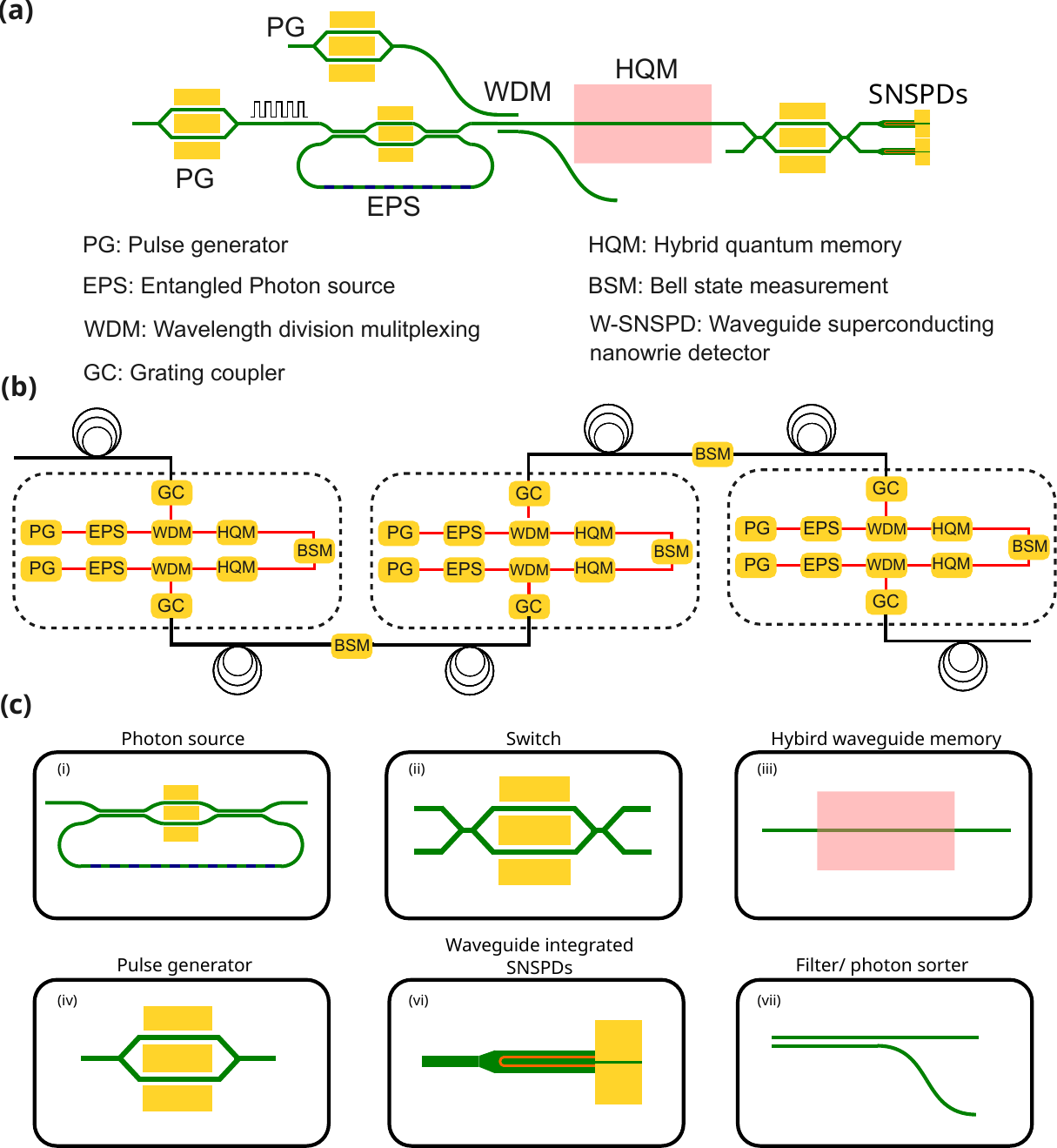}
    \caption{(a) Schematic of the unit cell of the proposed device architecture for on-chip quantum communication link. (b) Block diagram of three nodes, here each node is connected to the neighboring ones through optical fibers. (c) Simplified schematic of entangled photon sources (EPS), (ii) switch  (iii) hybrid quantum memory (HQM) (iv) pulse generator (PG), (v) superconducting nanowire single-photon detectors (SNSPDs), (vi) filter/ photon sorter).}
    \label{Fig_dev_arch}
\end{figure*}

Fig.\ref{Fig_dev_arch}(a) shows the proposed device architecture for each node, consisting of MZM modulators for pump-pulse generation for the entangled-photon source and an asymmetric periodically poled ring Mach-Zehnder interferometer  (PP-RMZI) for non-degenerate entangled photon-pair generation at the quantum-memory and the telecom wavelength. The non-degenerate photons can be filtered using simple directional couplers \cite{guo201670}. The memory photon is stored in a hybrid quantum memory, in which a YSO crystal is bonded to the LT PIC using oxide-oxide bonding. The memory regions consist of either straight or meandered waveguides, where a fraction of the optical mode is evanescently coupled to the YSO crystal. The memory can also be an impedance matched Fabry-perot (FP) cavity \cite{Afzelius2010ImpedanceMatchedCavity}. This coupling is possible because YSO has a lower refractive index than LT. The stored optical excitation is converted to a spin-wave excitation using control/read pulses and is eventually retrieved from the memory when required. The extracted photons from the memory are measured using waveguide-integrated superconducting nanowire single-photon detectors (W-SNSPDs), and Bell-state measurements are performed on-chip. The benefit of such a monolithic architecture is that only the telecom photons, which travel long distances through low-loss optical fibers, are coupled off-chip. In particular, the memory photons at shorter wavelengths remain on-chip, eliminating the need for efficient fiber-to-chip coupling at short wavelengths, which is challenging from a nanofabrication perspective.

For this paper, we will mostly focus on the entangled photon source and the hybrid memory as the are core components. Other components such as modulators for pulse generation, waveguide integrated SNSNDs, on-chip filters will be presented in our later experimental works. 

\begin{figure*} [!ht]
    \centering
    \includegraphics[width=0.90\textwidth]{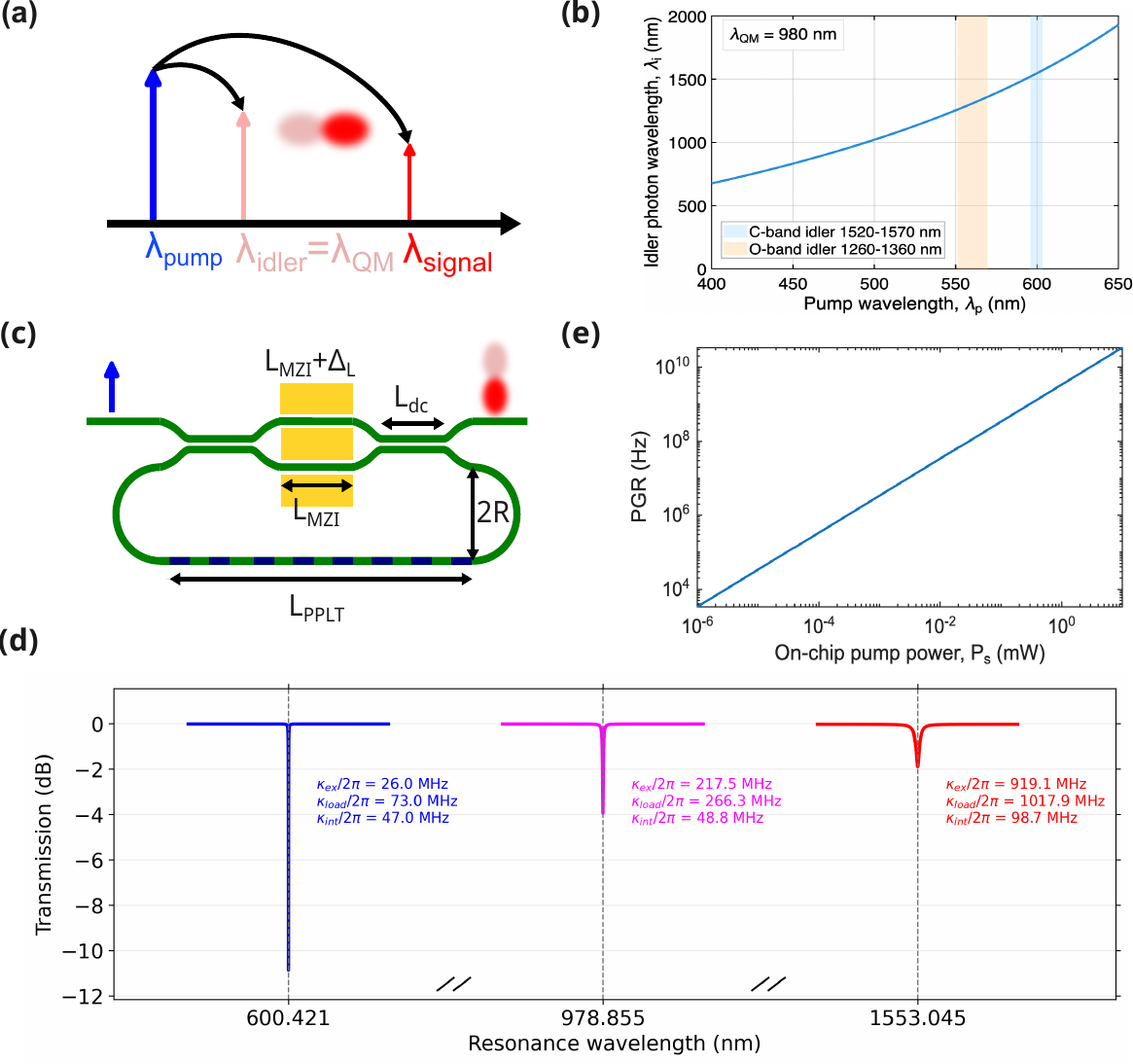}
    \caption{(a) Energy conversion diagram for the photon pair generation with spontaneous parametric down-conversion using three wave mixing process. (b) Idler photon wavelength as a function of the pump wavelength with signal photon fixed at $\mathrm{\lambda_{QM}=978.85\,nm}$, the shaded regions are represent the O-band and C-band. (c) Schematic of the PP-MRZI entangled photon pair source. (d) Simulated transmission from the device for the three modes with the corresponding coupling (loss) rates, external coupling rate, $\mathrm{\kappa_{ex}}$, intrinsic coupling rate, $\mathrm{\kappa_{int}}$, and loaded coupling rate, $\mathrm{\kappa_{load}}$. (e) Photon pair generation rate as a function of on-chip pump power. Here, the device parameters used in the simulation are, $\mathrm{L_{MZI}=1\,mm, L_{dc}=100\, \mu m, \Delta_{L}=5\,\mu m, R=100\,\mu m  }$.} 
    \label{Fig EPS}
\end{figure*}

\subsection{Entangled photon pair source}

AFC-based quantum memories require a high-rate entangled photon source capable of generating one photon at the quantum-memory transition wavelength $\lambda_{\mathrm{QM}}$ and the other at the telecom wavelength $\lambda_t$. Because of the recent development of hollow-core fibers with ultra-low loss near $1~\mu\mathrm{m}$, we can also assume that both photons can be generated at the quantum-memory wavelength, i.e., degenerate photon pairs at $\lambda_{\mathrm{QM}}$. However, degenerate photons are difficult to filter; hence, here we focus on non-degenerate photon-pair generation at the quantum-memory wavelength, $\lambda_{\mathrm{QM}}\sim980~\mathrm{nm}$, and at the C-band, $\lambda_t\sim1520$--$1570~\mathrm{nm}$. We have recently proposed PPLN (PPLT) ring--Mach--Zehnder interferometers (PP-RMZIs) as efficient photon-pair sources in the telecom band~\cite{kundu2025periodically} and also for quantum frequency conversion~\cite{kundu2025periodically1}. Here, we show how such devices can generate photon pairs in a non-degenerate fashion for the communication applications proposed here. Figure~\ref{Fig EPS}(a) shows the energy-conservation diagram for the photon-pair-generation process. In this process, one pump photon is converted into two lower-energy photons: one photon at the quantum-memory wavelength $\lambda_{\mathrm{QM}}$ and the other in the telecom band at $\lambda_t$. We consider a three-wave-mixing process based on the $\chi^{(2)}$ nonlinearity because $\chi^{(2)}$ is much stronger than $\chi^{(3)}$. In addition, lithium tantalate (LT) supports quasi-phase matching (QPM) through ferroelectric-domain engineering, which enables efficient nonlinear wavelength conversion. Figure~\ref{Fig EPS}(b) shows the energy-matched wavelengths when the signal photon is fixed at the quantum-memory wavelength, $\lambda_{\mathrm{QM}}\sim978.83~\mathrm{nm}$. To generate an idler photon in the O-band or C-band, the pump wavelength should be around $550~\mathrm{nm}$ or $600~\mathrm{nm}$, respectively, both of which fall within the transparency window of LT. Figure~\ref{Fig EPS}(c) shows a schematic of the proposed photon-pair-generation device. We propose an asymmetric ring--Mach--Zehnder interferometer (A-RMZI) that is triply resonant at the pump, signal, and idler wavelengths. In this design, the signal photon is generated at $\lambda_{\mathrm{QM}}$, and its bandwidth should be matched to the quantum-memory bandwidth. In contrast, the telecom idler photon does not have the same bandwidth constraint and should be over-coupled to maximize extraction and the heralding efficiency. The pump mode should be critically coupled to enable efficient energy conversion. Similar to our previous design, we consider only the straight section of the A-RMZI for the nonlinear interaction~\cite{kundu2025periodically,kundu2025periodically1}. We have recently shown that such RMZI resonators are electro-optically stable and that the coupling condition of the mode can be stably controlled from under-coupled to critically coupled to over-coupled~\cite{sayem2026high}. We used these devices as modulators for classical applications, where high-speed modulation is the key requirement. For photon-pair generation, however, we only need to set the coupling conditions for the three different modes with stability. We perform full-wave simulations together with rigorous numerical modeling. Our simulations include dispersion and material loss at the corresponding wavelengths. Details of the simulation method can be found in our previous works~\cite{kundu2025periodically,kundu2025periodically1}. In Fig.~\ref{Fig EPS}(d), we show the transmission spectra for the pump, signal, and idler modes after optimizing the device parameters and the bias voltage applied to the MZI. Here, we note that a wide parameter window is available to energy- and phase-match the three interacting modes. The parameters chosen for the current simulation are provided in the figure caption. One critical parameter assumed in the simulation is the propagation loss of the LT waveguide. At the pump and quantum-memory wavelengths, we assume $\alpha_{p,s}=0.1~\mathrm{dB/cm}$, while at the telecom wavelength, we assume $\alpha_i=0.2~\mathrm{dB/cm}$. This is a realistic propagation-loss value that has already been demonstrated in TFLN~\cite{desiatov2019visibleln} and partially in TFLT. Since the etching processes for LN and LT are similar, we can reasonably assume that such low loss is attainable with careful nanofabrication~\cite{zhu2026intrinsicLossTFLN,khalatpour2025roughnessLimitedTFLN}.In Fig.~\ref{Fig EPS}(e), we plot the photon-pair generation rate, $\mathrm{PGR}$, as a function of the pump power, $P_p$. As can be observed from Fig.~\ref{Fig EPS}(e), very high-rate entangled photons can be generated at very low pump power. The rate of generation will be limited by the minimum bandwidth of the generated photon pairs. A wide parameter window exists to match the wavelength and bandwidth of the generated photons to the memory transition wavelength and bandwidth. In this simulation, we show one example design.

\subsection{Hybrid quantum memory}

\begin{figure*}[!ht]
    \centering
    \includegraphics[width=1\textwidth]{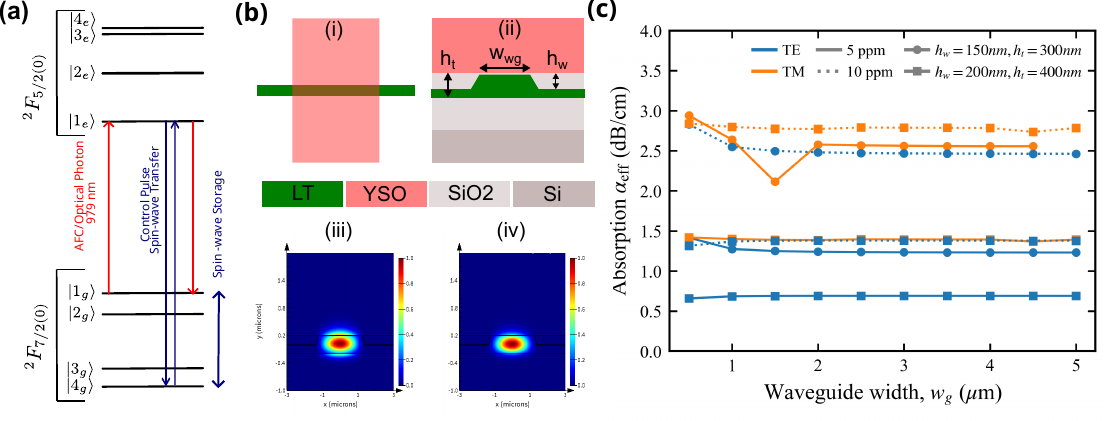}
    \caption{Hybrid YSO waveguide memory. (a) Simplified hyperfine-level structure of $^{171}\mathrm{Yb}^{3+}:\mathrm{Y}_{2}\mathrm{SiO}_{5}$, indicating the 979-nm AFC optical transition, the optical control transition, and the ground-state spin-wave coherence used for on-demand storage. (b) Cross sections of the two hybrid YSO--TFLT waveguide geometries considered here, together with representative fundamental TE-mode intensity profiles. (c) Effective Yb absorption coefficient, $\alpha_{\mathrm{eff}}$, as a function of waveguide width, $w_g$, for TE and TM modes, for Yb concentrations of 5 and 10 ppm and the waveguide geometries indicated in the legend. The simulated absorption remains large over a broad range of waveguide widths, enabling strong light--matter interaction in the proposed integrated memory.}
    \label{HM}
\end{figure*}


\begin{figure*}[!ht]
    \centering
    \includegraphics[width=1\textwidth]{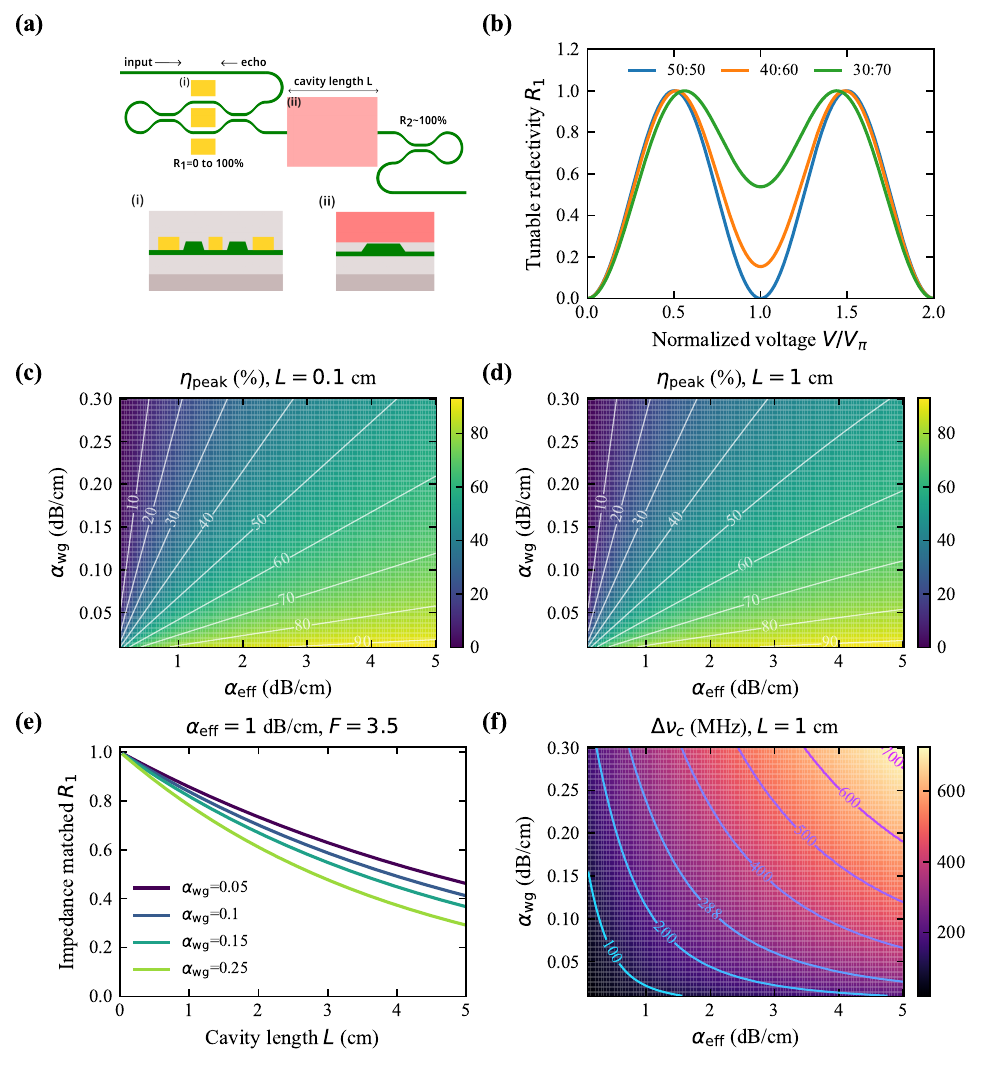}
    \caption{Tunable cavity design and performance of the integrated AFC quantum memory. (a) Schematic of the proposed cavity architecture, where a tunable Mach--Zehnder-interferometer-based input coupler controls the effective front-mirror reflectivity $R_1$, while the back reflector is taken to be $R_2=100\%$. Insets show the corresponding cross-sectional device geometries. (b) Tunable reflectivity $R_1$ as a function of normalized drive voltage $V/V_{\pi}$ for different directional-coupler splitting ratios. (c,d) Peak AFC memory efficiency $\eta_{\mathrm{peak}}$ as a function of effective AFC absorption $\alpha_{\mathrm{eff}}$ and waveguide propagation loss $\alpha_{\mathrm{wg}}$ for cavity lengths $L=0.1~\mathrm{cm}$ and $L=1~\mathrm{cm}$, respectively. (e) Impedance-matched input-mirror reflectivity $R_1$ as a function of cavity length for $\alpha_{\mathrm{eff}}=1~\mathrm{dB/cm}$, $F=3.5$, and several values of $\alpha_{\mathrm{wg}}$. (f) Corresponding cavity linewidth $\Delta\nu_c$ for $L=1~\mathrm{cm}$ as a function of $\alpha_{\mathrm{eff}}$ and $\alpha_{\mathrm{wg}}$.}
    \label{HM}
\end{figure*}

\subsubsection{Device architecture}

In this section, we explore the hybrid $\Yb{:}\YSO$-TFLT quantum memory device structure. A $\Yb{:}\YSO$ crystal is bonded on top of a TFLT waveguide,  as shown in Fig.\ref{HM}(a), so that the guided mode is confined by the TFLT waveguide while its evanescent tail overlaps with the $\Yb$ ions in the $\YSO$ crystal.  The hybrid memory may consist of a straight waveguide or a meander waveguide, read out in a single-pass forward or backward readout scheme; or in an impedance-matched cavity architecture~\cite{Afzelius2010ImpedanceMatchedCavity}, where the $\YSO$ crystal is bonded to the cavity section of a one-sided Fabry--Perot (FP) cavity with a tunable input mirror~\cite{sayem2025tunable}, as shown in Fig.\ref{HM}. Figs.\ref{HM}(b)(i) and \ref{HM}(b)(ii) show the cross-section of the TFLT waveguide with a thin oxide layer on top after polishing and the cross-section of the structure where a $\YSO$ crystal is bonded on top of the TFLT waveguide, respectively. A similar bonding technique has been successfully demonstrated for $\YSO$ and TFLN waveguides/resonators~\cite{yang2021photonic}, and with current state-of-the-art fabrication technology, this process is straightforward. For this hybrid quantum memory structure to work, the absorption loss must be dominated by the $\Yb$ ions in the $\YSO$ crystal rather than by propagation loss in the TFLT waveguides. TFLN and TFLT waveguides have shown ultra-low loss at both telecom wavelengths, specifically the C-band, and visible wavelengths~\cite{zhang2017monolithic,desiatov2019visibleln}. We assume the propagation loss of the TFLT waveguides to be in the range of $\mathrm{0.1\,dB/cm}$, which is close to what has been achieved near the C-band~\cite{zhang2017monolithic,desiatov2019visibleln}. Operation near $\lambda=\mathrm{1\,\mu m}$ offers even tighter mode confinement, especially with wider waveguides, and is therefore less affected by scattering loss due to surface roughness, a major contributor to propagation loss~\cite{zhu2026intrinsicLossTFLN,khalatpour2025roughnessLimitedTFLN}.

In Fig.\ref{HM}(b), we show the optical mode profiles for the fundamental TE and TM modes, demonstrating the waveguide mode overlap with the crystal. For our simulation, we assume a spin-polarized, integrated absorption of
$\alpha_{\YSO}=1.6~\mathrm{cm^{-1}}$ at $5$~ppm doping concentration~\cite{Businger2022YbYSO} with a linear increase in absorption with doping concentration. In Fig.\ref{HM}(c), we plot the absorption in the evanescently coupled $\Yb{:}\YSO$ as a function of TFLT waveguide width, $w_{\mathrm{g}}$, for different film thicknesses and etch depths. We discuss our modeling and simulation of absorption in the supplementary materials. From Fig.\ref{HM}(c), we observe that a high absorption rate can be achieved. For example, for a film thickness of $h_{\mathrm{t}}=\mathrm{360\,nm}$ and an etch depth of $h_{\mathrm{w}}=\mathrm{180\,nm}$, absorption by the YSO crystal reaches $\mathrm{0.9\,dB/cm}$, which is significantly higher than the typical waveguide loss achieved in these material platforms, typically around $\mathrm{0.1\,dB/cm}$--$\mathrm{0.2\,dB/cm}$. This indicates that the YSO-TFLT hybrid memory does not require complicated waveguide tapering or geometry optimization, unlike diamond integration with TFLN/TFLT. The simulation parameters are provided in the caption of Fig.\ref{HM}. This loss rate is also high enough for impedance-matched cavities to achieve near-unity memory efficiency, as shown in our later analysis. 

\subsubsection{Analysis of AFC efficiency}
\label{sec:cavity}
An atomic frequency comb (AFC) quantum memory stores an input photon as a collective coherence distributed over a periodic comb of absorbing teeth tailored on an inhomogeneously broadened optical transition, and re-emits it as an echo at the re-phasing time, $t_s=\frac{1}{\Delta}$, where $\Delta$ is the comb tooth spacing. The optical coherence can be transferred to spin coherence with optical control pulses, controlled by microwave pulses applied through a microwave resonator such as superconducting loop-gap resonator, enabling long-lived spin-wave storage and on-demand retrieval. However, in this work, we focus our analysis on optical AFC storage and leave the investigation of spin-wave storage in the hybrid device architecture for future work. We primarily concentrate on the impedance-matched cavity architecture in the main article. 

since single-pass forward retrieval efficiency is fundamentally limited to $\approx54\%$ by re-absorption of the emitted echo, while near-unity backward retrieval requires an additional phase-matching operation and associated control fields~\cite{Afzelius2009AFC}. An impedance-matched cavity provides an alternative route to near-unity efficiency without requiring backward retrieval~\cite{Afzelius2010ImpedanceMatchedCavity}; moreover, in the present integrated platform, the required impedance-matching condition can be readily adjusted using an electro-optically tunable input mirror~\cite{sayem2025tunable}. Additionally, unlike bulk crystal implementations, in the device architecture, waveguide propagation loss acts over the very length that stores and re-emits the echo, and it must be carried through the efficiency analysis from the outset.  We therefore derive the cavity echo efficiency in the presence of finite propagation and background losses and identify the parameter regime that maximizes the attainable efficiency. For completeness, we provide detailed derivations for the single-pass forward and backward retrieval and impedance-matching cavity geometries, along with the corresponding optimization of their AFC and device parameters, in the Supplementary Information.

Three absorption coefficients act on the guided mode. The comb-averaged useful absorption, $\alpha\equiv\aeff/F$, where $F$ is the comb finesse, is the only channel that stores recoverable atomic coherence. It therefore defines the useful optical depth, $\dtil\equiv\alpha L$. In contrast, the waveguide propagation loss $\awg$ and the inter-tooth background absorption $\azero$ are broadband loss channels. Both attenuate the optical field wherever it propagates, including on the comb teeth and in the spectral gaps, but neither contributes to the stored coherence. Since these two broadband loss mechanisms are geometrically indistinguishable, we combine them into a total broadband loss coefficient, $\aell\equiv\awg+\azero$. The total single-pass optical depth is then
\begin{equation}
s \equiv \dtil+\dpar+\dzero=(\alpha+\aell)L,
\label{eq:s}
\end{equation}
where $\dpar\equiv\awg L$ and $\dzero\equiv\azero L$. The comb dephasing factor $D(F)$ multiplies the emitted intensity and does not enter the spatial integral. For square comb teeth, $D(F)=\mathrm{sinc}^{2}(\pi/F)$. An intensity optical depth $d$ attenuates the field amplitude by $e^{-d/2}$. Because the AFC echo is obtained from a coherent sum of the fields emitted at different positions, we therefore work with field amplitudes throughout the derivation and square the final result to obtain the efficiency. For the impedance-matched cavity geometry, the bonded waveguide forms a one-sided Fabry--P'erot resonator with an input coupling mirror of intensity reflectivity $R_1$ and a near-unity back mirror of reflectivity $R_2$. The guided mode is absorbed and subsequently re-emitted over the interaction length $L$. A single pass through the waveguide multiplies the field amplitude by $e^{-(\alpha+\aell)L/2}$, while one complete cavity round trip consists of two full passes through the waveguide together with reflection from the back mirror. The corresponding round-trip amplitude factor is therefore $A=\sqrt{R_2},e^{-(\alpha+\aell)L}$. Following Afzelius and Simon~\cite{Afzelius2010ImpedanceMatchedCavity}, when the inhomogeneous linewidth is much larger than the homogeneous linewidth, $\gamma_i\gg\gamma_h$, the atomic ensemble can be treated as an effective intracavity loss. Impedance matching is then obtained when the input-coupling loss equals the internal round-trip loss,
\begin{equation}
R_1=R_2,e^{-2(\alpha+\aell)L}
=R_2\exp\Big[-2\Big(\tfrac{\aeff}{F}+\awg+\azero\Big)L\Big].
\label{eq:match}
\end{equation}

The echo field is obtained by coherently summing the contributions from all positions $z\in[0,L]$ along the waveguide. A differential layer $dz$ is reached by the cavity-enhanced input field, which is proportional to $\sqrt{T_1},e^{-(\alpha+\aell)z/2}/(1-q)$, where $T_1=1-R_1$ and $q=\sqrt{R_1R_2},e^{-(\alpha+\aell)L}$ is the cavity build-up ratio. The layer stores coherence with useful weight $\alpha,dz$.

The echo emitted by this layer then propagates from $z$ to the back mirror, is reflected, and returns toward the output coupler. Including the second cavity build-up, this contribution is proportional to $e^{-(\alpha+\aell)(L-z)/2}e^{-(\alpha+\aell)L/2}\sqrt{R_2T_1}/(1-q)$. The standing-wave field gives an additional factor of $2$ because absorption occurs for both propagation directions. Importantly, the $z$-dependent attenuation factors cancel when the input and echo propagation factors are multiplied. Their product reduces to the constant $e^{-(\alpha+\aell)L}$. The spatial integral therefore reduces simply to $\int_0^L\alpha,dz=\alpha L$, giving
\begin{equation}
\sqrt{\eta}
=\frac{2,(\alpha L),e^{-(\alpha+\aell)L},T_1\sqrt{R_2},\sqrt{D(F)}}
{\big(1-\sqrt{R_1R_2},e^{-(\alpha+\aell)L}\big)^{2}}.
\label{eq:sqrteta}
\end{equation}

This expression reproduces Eq.~(14) of Ref.~\cite{Afzelius2009AFC}. The distinction between useful and parasitic absorption is explicit: the useful coefficient $\alpha$ appears only through the factor $\alpha L$, because only the comb contributes to the recoverable coherence, whereas the total coefficient $\alpha+\aell$ determines all propagation losses. 
Under the impedance-matching condition in Eq.~\eqref{eq:match}, the denominator reduces to $T_1$. For an ideal back mirror, $R_2=1$. Using $1-e^{-2x}=2e^{-x}\sinh x$, with $x=(\alpha+\aell)L$, the matched-cavity efficiency can then be written directly in terms of the physical absorption coefficients as
\begin{equation}
\eta_{\mathrm{match}}
=\left[
\frac{(\aeff/F)L}
{\sinh\!\big[(\aeff/F+\awg+\azero)L\big]}
\right]^{2}
D(F),
\label{eq:eta-exact}
\end{equation}
which is exact for any interaction length $L$.

\subsubsection{Optimal parameters for efficiency and bandwidth}
\label{sec:criteria}

The exact impedance-matched efficiency can be written in a form that makes the relevant device tradeoffs explicit,
\begin{equation}
\eta_{\mathrm{match}}
=
\frac{D(F)}{(1+bF)^2}
\left[\frac{s}{\sinh s}\right]^2,
\qquad
b\equiv\frac{\aell}{\aeff},
\label{eq:eta_design}
\end{equation}
where the three factors describe AFC dephasing, parasitic loss relative to useful absorption loss, and the finite-length penalty, respectively.

An important consequence for the hybrid architecture is that the peak efficiency is controlled by the ratio
$b=\aell/\aeff$, rather than by optical depth alone. Since
$[s/\sinh s]^2\leq1$, the matched efficiency satisfies
\begin{equation}
\eta_{\mathrm{match}}
\leq
\frac{D(F)}{(1+bF)^2}.
\label{eq:eta_ceiling}
\end{equation}
Thus, broadband attenuation of the guided mode imposes an intrinsic efficiency ceiling even for a perfectly impedance-matched cavity. The quantity $b$ is independent of device length and is determined primarily by the broadband parasitic loss $\aell$ and the useful rare-earth absorption $\aeff$. This is distinct from the usual free-space picture, in which optical depth is often the primary figure of merit. 
Figures~\ref{fig:HM2}(b,c) illustrate this behavior. Increasing $\aeff$ or reducing $\aell$ raises the achievable efficiency, whereas increasing the interaction length enhances the finite-length penalty. Nevertheless, realistic device parameters can support maximum zero-delay efficiencies exceeding $80\%$. The AFC finesse introduces a second tradeoff. Increasing $F$ suppresses AFC dephasing and therefore increases $D(F)$, but it also reduces the useful comb-averaged absorption from $\aeff$ to $\aeff/F$. As a result, parasitic loss becomes more important through the factor $(1+bF)^{-2}$, while the effective absorption density available to the cavity is reduced. The efficiency therefore exhibits an optimum finesse, $F_{\mathrm{opt}}$, for each set of device parameters. The Supplementary Information provides the full optimization with respect to both $F$ and $L$, including the dependence of $F_{\mathrm{opt}}$ on propagation loss and absorption.

For a chosen $F$ and $L$, the impedance-matching condition fixes the required input-mirror reflectivity,
\begin{equation}
R_1=
R_2\exp\left[
-2\left(\frac{\aeff}{F}+\aell\right)L
\right].
\label{eq:R1_design}
\end{equation}
As shown in Fig.~\ref{fig:HM2}(d), the matched value of $R_1$ decreases as either the cavity length or the internal attenuation increases. In the hybrid architecture, however, the input mirror does not need to be fixed during fabrication. Instead, its effective reflectivity can be actively tuned electro-optically \cite{sayem2025tunable}. Figure~\ref{fig:HM2}(f) shows the corresponding voltage-dependent tuning of the mirror reflectivity, which enables the cavity to be impedance matched after the AFC is prepared and the internal device loss is characterized.

The AFC bandwidth introduces an additional constraint because the prepared comb must lie within the impedance-matched cavity resonance,
$\GAFC\lesssim\Delta\nu_c$. Under the matching condition, the cavity linewidth is
\begin{equation}
\Delta\nu_c
=
\frac{c}{\pi n_g L}
\sinh\left(\frac{s}{2}\right)
=
\frac{c}{2\pi n_g}
\left(\frac{\aeff}{F}+\aell\right)
\frac{\sinh(s/2)}{s/2}.
\label{eq:cavity_linewidth}
\end{equation}
This expression shows that the linewidth is governed primarily by the total absorption density $\aeff/F+\aell$, rather than by cavity length alone. Shortening the device increases the free spectral range, but impedance matching simultaneously requires a more reflective input mirror, causing the two effects to partially compensate. Figure~\ref{fig:HM2}(e) shows the resulting linewidth over the relevant range of absorption and propagation loss. The finesse that maximizes efficiency is therefore not necessarily the optimum choice when a larger AFC bandwidth is required. Increasing $F$ reduces AFC dephasing, but it also lowers $\aeff/F$ and consequently narrows the impedance-matched cavity resonance. Table~\ref{tab:criteria} illustrates this tradeoff for $\aeff=2~\mathrm{dB/cm}$, $\aell=0.10~\mathrm{dB/cm}$, $n_g=2.2$, and square AFC teeth, corresponding to $b=0.05$. The efficiency reaches its maximum near $F_{\mathrm{opt}}=4.36$, whereas a moderately smaller finesse provides a substantially wider cavity linewidth with only a modest reduction in efficiency. The appropriate operating finesse therefore depends on whether peak storage efficiency or larger AFC bandwidth is the primary requirement. These results suggest two practical design guidelines. First, performance is improved most effectively by reducing the ratio $b=\aell/\aeff$, through suppression of broadband waveguide loss and enhancement of useful rare-earth absorption, rather than by simply increasing the interaction length. Second, the AFC finesse should be chosen near the efficiency optimum, or slightly below it when additional cavity bandwidth is required.

\begin{table}[t]
    \centering
    \caption{Matched AFC efficiency and cavity linewidth for
    $\aeff=2~\mathrm{dB/cm}$, $\aell=0.10~\mathrm{dB/cm}$,
    $n_g=2.2$, and square comb teeth ($b=0.05$).}
    \label{tab:criteria}
    \begin{tabular}{c c c}
        \hline\hline
        $F$ & $\eta$ & $\Delta\nu_c$ (MHz) \\
        \hline
        2.5 & 45.3\% & 449 \\
        3.0 & 51.7\% & 383 \\
        $F_{\mathrm{opt}}=4.36$ & 56.5\% & 279 \\
        5.0 & 56.0\% & 250 \\
        8.0 & 48.5\% & 175 \\
        \hline\hline
    \end{tabular}
\end{table}

\begin{figure*} [!ht]
    \centering
    \includegraphics[width=1\linewidth,height=0.70\textheight,keepaspectratio]{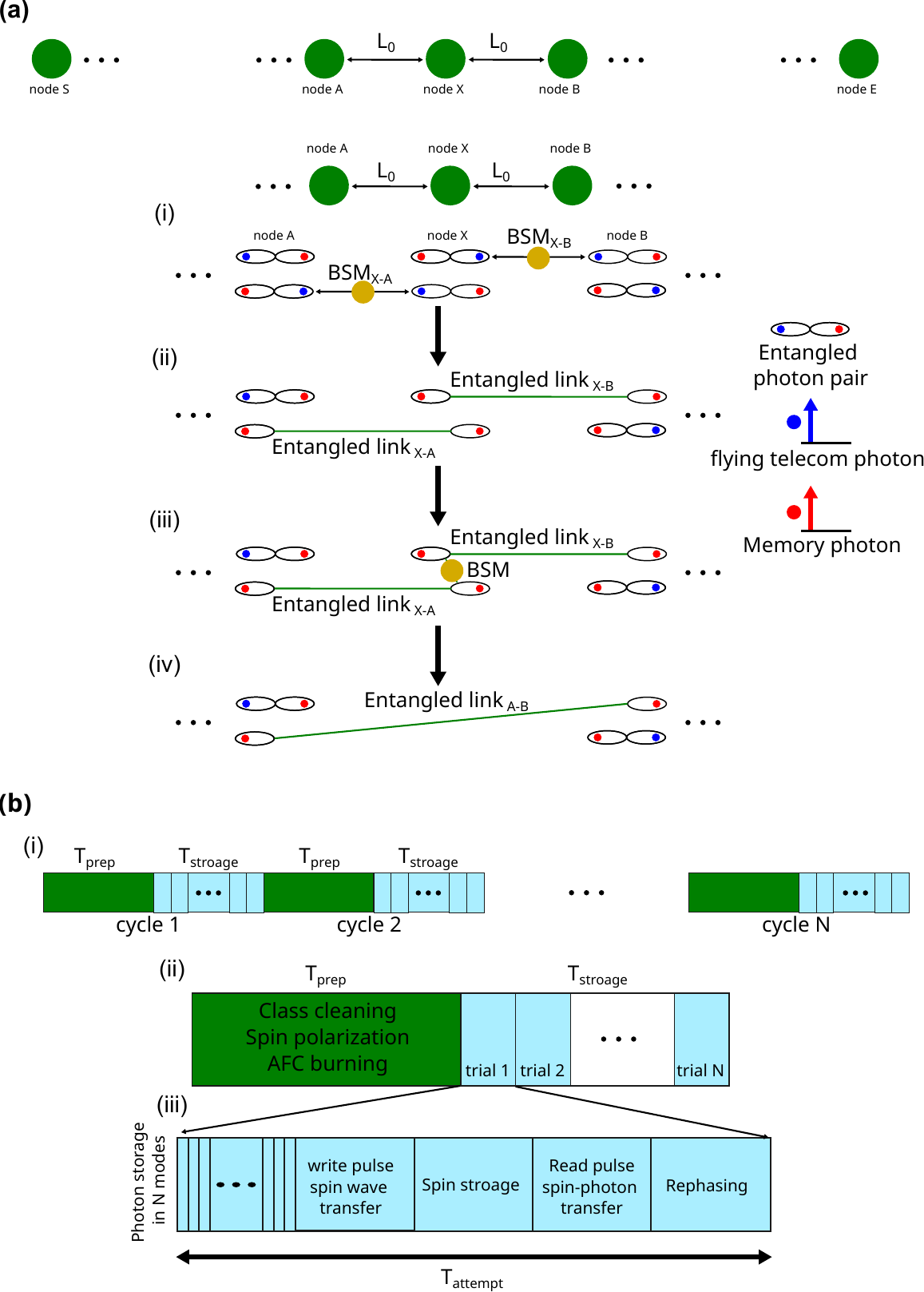}
    \caption{Schematic of the end to end entanglement link. Each node is connected to the neighboring ones through the flying photons from EPS from each node. (i)-(iv) Example entanglement link between node A and node B through node X. Successful BSMs of the telecom photons between node A-X and X-B establish the entanglement between the memory photons between node A-X and X-B. BSM of the extracted memory photons entangle the memory photons at node A and B establishing the entanglement between node A and B. (b) Timing sequence for the AFC-based quantum memory operation. (i) The memory is operated periodically over multiple cycles, where each cycle consists of a preparation stage, $T_{\mathrm{prep}}$, followed by a storage stage, $T_{\mathrm{storage}}$. (ii) During $T_{\mathrm{prep}}$, the memory is prepared through class cleaning and AFC comb burning. The prepared comb can then be used for multiple storage trials within $T_{\mathrm{storage}}$. (iii) Each storage attempt, with duration $T_{\mathrm{attempt}}$, consists of photon storage in $N$ multiplexed modes, spin-wave conversion using a write pulse, spin storage, readout using a read pulse, and rephasing before the next attempt.
 }
    \label{RATE1}    
\end{figure*}

\section{Rate and fidelity analysis for the memory-assisted link}
\label{sec:rate}

In this section, we evaluate the elementary-link success probability, the single-link entanglement-generation rate and fidelity, and the end-to-end rate for repeater chains. Fig.~\ref{RATE1} shows the link architecture. For each elementary link, the two $1550$-nm photons generated by neighboring source--memory units are sent to a midpoint Bell-state measurement (BSM), while the corresponding $\sim980$-nm photons are stored in the two AFC memories. A successful midpoint BSM therefore heralds an entangled pair of memory excitations. Neighboring elementary links are then connected by recalling the stored photons and performing a second, local BSM at the shared node. We consider one hybrid memory per link direction and chains containing up to four equal elementary links. A larger on-chip memory bank (spatially multiplexed) is expected to reduce source blocking and memory waiting times, but is not included in the present calculation.

The model keeps loss and conditional-state errors separate. Fiber attenuation, finite storage efficiency, and imperfect recall primarily reduce the probability of a successful event. Multipair generation, detector noise, and residual phase errors can instead produce an accepted event with reduced Bell fidelity. This separation follows the standard treatment of memory-assisted
quantum links and allows the rate and conditional fidelity to be evaluated with the same physical parameters~\cite{sangouard2011,panayi2014memory}.

\begin{figure*} [!ht]
    \centering
    \includegraphics[width=1\textwidth]{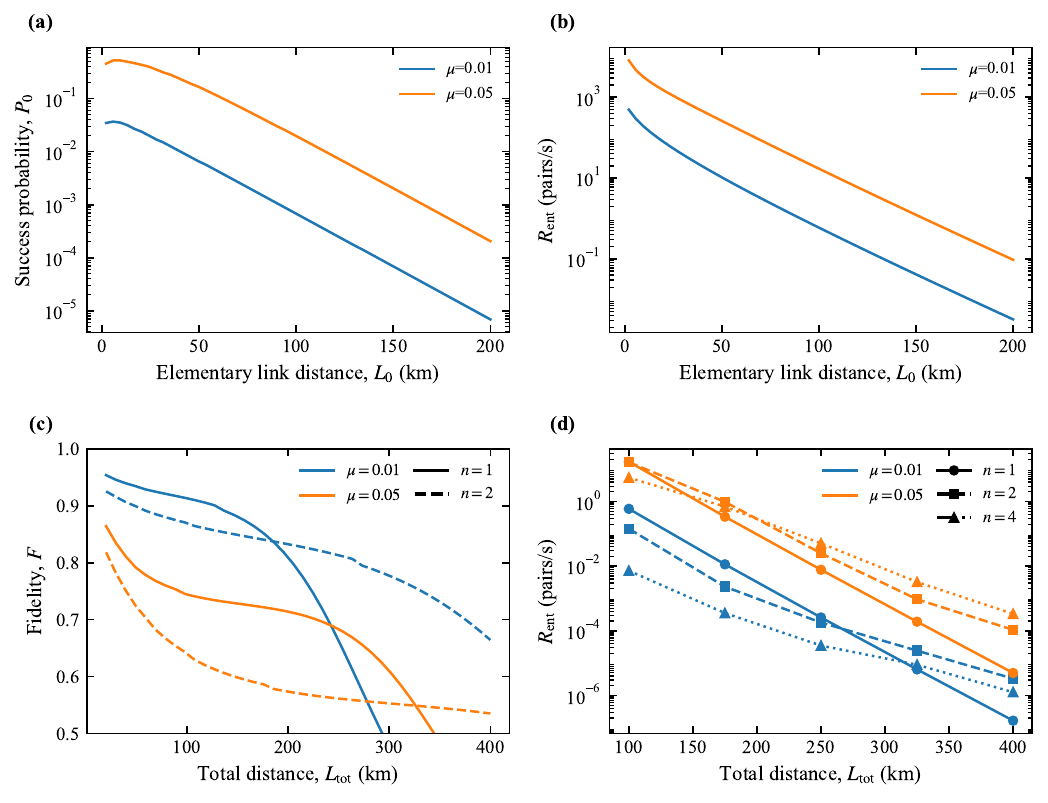}
   \caption{ Rate and fidelity of the memory-assisted link. (a)~Elementary-link train-success probability $P_0$ and (b)~single-link entanglement-generation rate $R_0$ as a function of elementary-link length $L_0$, for source mean pair number $\mu=0.01$, $0.05$ at $f_{\mathrm{clk}}=100$~MHz. At each distance, the temporal-mode number $N$ is chosen to maximize $R_0$ while including the finite AFC-coherence penalty. (c)~Conditional single-link Bell-state fidelity $F$ for the same two source brightnesses. The reduction in fidelity at higher $\mu$ results from the increased contribution of multipair sectors to accepted BSM events. (d)~End-to-end entanglement-generation rate $R_{\mathrm{ent}}$ for chains containing $n=1$, $2$ and $4$ equal elementary links. Successful segments are stored while neighboring links continue to generate entanglement, and a failed local BSM regenerates only the two consumed child segments} 

    \label{RATE2}    
\end{figure*}

\begin{table*}[t]
\centering
\caption {Representative parameters used in the rate and fidelity calculations of Sec.~V. }
\label{tab:network_model_parameters}

\small
\setlength{\tabcolsep}{3pt}
\renewcommand{\arraystretch}{1.15}

\begin{tabular*}{\textwidth}{@{\extracolsep{\fill}} l c c l}
\toprule

\parbox[t]{0.18\textwidth}{\raggedright\textbf{Parameter}\strut}
&
\textbf{Symbol}
&
\textbf{Value}
&
\parbox[t]{0.46\textwidth}{\raggedright\textbf{Description}\strut}
\\

\midrule

\parbox[t]{0.18\textwidth}{\raggedright Source clock rate\strut}
&
$f_{\mathrm{clk}}$
&
$100~\mathrm{MHz}$
&
\parbox[t]{0.46\textwidth}{\raggedright
Effective temporal-mode repetition rate of the cavity-enhanced photon-pair
source, corresponding to
$T_{\mathrm{mode}}=1/f_{\mathrm{clk}}=10~\mathrm{ns}$.
\strut}
\\

\parbox[t]{0.18\textwidth}{\raggedright Mean generated pair number\strut}
&
$\mu$
&
$0.01,\;0.05,\;$
&
\parbox[t]{0.46\textwidth}{\raggedright
Mean photon-pair number per effective temporal mode for each source.
Single-mode two-mode-squeezed-vacuum statistics are assumed.
\strut}
\\

\parbox[t]{0.18\textwidth}{\raggedright Source coherence\strut}
&
$C_{\mathrm{src}}$
&
$0.99$
&
\parbox[t]{0.46\textwidth}{\raggedright
Intrinsic coherence of each generated energy-time-entangled photon pair.
\strut}
\\

\parbox[t]{0.18\textwidth}{\raggedright Maximum source pair number\strut}
&
$n_{\max}$
&
$6$
&
\parbox[t]{0.46\textwidth}{\raggedright
Highest TMSV pair-number sector retained in the numerical summation.
\strut}
\\

\parbox[t]{0.18\textwidth}{\raggedright Number of AFC temporal modes\strut}
&
$N$
&
Optimized
&
\parbox[t]{0.46\textwidth}{\raggedright
Optimized independently at each elementary-link length over
$100\leq N\leq2\times10^{4}$.
\strut}
\\

\parbox[t]{0.18\textwidth}{\raggedright Fiber attenuation\strut}
&
$\alpha_{\mathrm{fib}}$
&
$0.20~\mathrm{dB/km}$
&
\parbox[t]{0.46\textwidth}{\raggedright
Attenuation of the $1550$-nm telecom channel. Each photon propagates over
$L_0/2$ to the midpoint BSM.
\strut}
\\

\parbox[t]{0.18\textwidth}{\raggedright Fiber group velocity\strut}
&
$v_{\mathrm f}$
&
$2.0\times10^{8}~\mathrm{m/s}$
&
\parbox[t]{0.46\textwidth}{\raggedright
Used to calculate photon-propagation and heralding latency.
\strut}
\\

\parbox[t]{0.18\textwidth}{\raggedright Telecom coupling efficiency\strut}
&
$\eta_{\mathrm{c}}$
&
$0.95$
&
\parbox[t]{0.46\textwidth}{\raggedright
Coupling and routing efficiency for each telecom photon.
\strut}
\\

\parbox[t]{0.18\textwidth}{\raggedright Telecom detector efficiency\strut}
&
$\eta_{\mathrm{d}}$
&
$0.95$
&
\parbox[t]{0.46\textwidth}{\raggedright
Detection efficiency of the telecom photon-number-resolving detector.
\strut}
\\

\parbox[t]{0.18\textwidth}{\raggedright Midpoint BSM efficiency\strut}
&
$\eta_{\mathrm{BSM}}^{(1550)}$
&
$0.50$
&
\parbox[t]{0.46\textwidth}{\raggedright
Intrinsic success probability of the linear-optical midpoint Bell-state
measurement.
\strut}
\\

\parbox[t]{0.18\textwidth}{\raggedright Telecom HOM visibility\strut}
&
$V_{\mathrm{HOM}}^{(1550)}$
&
$0.99$
&
\parbox[t]{0.46\textwidth}{\raggedright
Two-photon interference visibility at the midpoint BSM.
\strut}
\\

\parbox[t]{0.18\textwidth}{\raggedright Zero-delay AFC efficiency\strut}
&
$\eta_{\mathrm{AFC},0}$
&
$0.8$
&
\parbox[t]{0.46\textwidth}{\raggedright
Representative impedance-matched AFC efficiency used as the zero-delay
memory efficiency.
\strut}
\\

\parbox[t]{0.18\textwidth}{\raggedright Effective AFC coherence time\strut}
&
$T_{2,\mathrm{AFC}}$
&
$307~\mu\mathrm{s}$
&
\parbox[t]{0.46\textwidth}{\raggedright
Used in
$\eta_{\mathrm{AFC}}
=\eta_{\mathrm{AFC},0}
\exp[-4\tau_{\mathrm{AFC}}/T_{2,\mathrm{AFC}}]$.
\strut}
\\

\parbox[t]{0.18\textwidth}{\raggedright Write-control efficiency\strut}
&
$\eta_{\pi,\mathrm w}$
&
$0.9$
&
\parbox[t]{0.46\textwidth}{\raggedright
Efficiency of the optical write-control transfer.
\strut}
\\

\parbox[t]{0.18\textwidth}{\raggedright Read-control efficiency\strut}
&
$\eta_{\pi,\mathrm r}$
&
$0.9$
&
\parbox[t]{0.46\textwidth}{\raggedright
Efficiency of the optical read-control transfer.
\strut}
\\

\parbox[t]{0.18\textwidth}{\raggedright Spin coherence time\strut}
&
$T_{2,\mathrm{spin}}$
&
$10~\mathrm{ms}$
&
\parbox[t]{0.46\textwidth}{\raggedright
Baseline spin-coherence timescale used in the present calculation.
\strut}
\\

\parbox[t]{0.18\textwidth}{\raggedright Spin-decay exponent\strut}
&
$\beta$
&
$1$
&
\parbox[t]{0.46\textwidth}{\raggedright
Exponent used in the spin-coherence decay model.
\strut}
\\

\parbox[t]{0.18\textwidth}{\raggedright Memory preparation time\strut}
&
$T_{\mathrm{prep}}$
&
$0.50~\mathrm{s}$
&
\parbox[t]{0.46\textwidth}{\raggedright
Time assigned to class cleaning, spin polarization, and AFC preparation.
\strut}
\\

\parbox[t]{0.18\textwidth}{\raggedright Prepared-comb active time\strut}
&
$T_{\mathrm{storage}}$
&
$0.50~\mathrm{s}$
&
\parbox[t]{0.46\textwidth}{\raggedright
Time over which the prepared AFC is assumed to support repeated trials.
\strut}
\\

\parbox[t]{0.18\textwidth}{\raggedright Preparation duty cycle\strut}
&
$D$
&
$0.50$
&
\parbox[t]{0.46\textwidth}{\raggedright
Calculated as
$D=T_{\mathrm{storage}}/(T_{\mathrm{prep}}+T_{\mathrm{storage}})$.
\strut}
\\

\parbox[t]{0.18\textwidth}{\raggedright Local routing efficiency\strut}
&
$\eta_{\mathrm{route}}$
&
$0.95$
&
\parbox[t]{0.46\textwidth}{\raggedright
On-chip transmission and routing efficiency for each recalled $980$-nm photon.
\strut}
\\

\parbox[t]{0.18\textwidth}{\raggedright Memory-photon detector efficiency\strut}
&
$\eta_{\mathrm{d,m}}$
&
$0.95$
&
\parbox[t]{0.46\textwidth}{\raggedright
PNR detector efficiency for each recalled memory photon.
\strut}
\\

\parbox[t]{0.18\textwidth}{\raggedright Local BSM efficiency\strut}
&
$\eta_{\mathrm{BSM}}^{(980)}$
&
$0.50$
&
\parbox[t]{0.46\textwidth}{\raggedright
Intrinsic success probability of the linear-optical BSM between two recalled
memory photons.
\strut}
\\

\parbox[t]{0.18\textwidth}{\raggedright Processing latency\strut}
&
$T_{\mathrm{proc}}$
&
$1~\mu\mathrm{s}$
&
\parbox[t]{0.46\textwidth}{\raggedright
Additional local electronic and control-processing time per attempt.
\strut}
\\

\parbox[t]{0.18\textwidth}{\raggedright False-sector Bell fidelity\strut}
&
$F_{\mathrm{false}}$
&
$1/4$
&
\parbox[t]{0.46\textwidth}{\raggedright
Conservative Bell-state fidelity assigned to accepted multipair sectors
other than the desired $(1,1)$ source sector.
\strut}
\\

\parbox[t]{0.18\textwidth}{\raggedright Number of elementary links\strut}
&
$n$
&
$1,\;2,\;4$
&
\parbox[t]{0.46\textwidth}{\raggedright
Nested repeater chains considered in the end-to-end rate calculation.
\strut}
\\

\bottomrule
\end{tabular*}
\end{table*}

\subsection{Memory efficiency and elementary-link probability}

For an AFC with comb spacing $\Delta$, the programmed optical rephasing delay is $\tau_{\mathrm{AFC}}=1/\Delta$. The corresponding optical AFC efficiency is written as
\begin{equation}
\eta_{\mathrm{AFC}}(\tau_{\mathrm{AFC}})
=\eta_0\exp\!\left(-\frac{4\tau_{\mathrm{AFC}}}{T_{2,\mathrm{AFC}}}\right),
\label{eq:etaAFC_main}
\end{equation}
where $\eta_0$ is the short-delay efficiency obtained from the cavity-memory analysis of Sec.~IV. The AFC delay must accommodate the complete input train and the write-control pulse before the first echo. We therefore use
\begin{equation}
\tau_{\mathrm{AFC}}\simeq NT_{\mathrm{mode}}+T_{\mathrm{ctrl}}+T_{\mathrm{guard}},
\label{eq:tauAFC_main}
\end{equation}
with $T_{\mathrm{mode}}=1/f_{\mathrm{clk}}$. Increasing the number of temporal modes therefore provides more entanglement-generation attempts, but also increases the AFC delay and lowers the storage efficiency. The optimal $N$ is chosen numerically for each elementary-link length rather than fixed at the maximum available multimode capacity.

For spin-wave operation, the complete store-and-retrieve efficiency is
\begin{equation}
\eta_{\mathrm{mem}}(\tau_{\mathrm{AFC}},T_s)
=\eta_{\mathrm{AFC}}(\tau_{\mathrm{AFC}})
\eta_{\pi,w}\eta_{\pi,r}\eta_{\mathrm{spin}}(T_s),
\label{eq:etamem_main}
\end{equation}
where $\eta_{\pi,w}$ and $\eta_{\pi,r}$ are the optical-to-spin and spin-to-optical transfer efficiencies, respectively, and
$\eta_{\mathrm{spin}}(T_s)$ accounts for the loss of collective spin-wave rephasing during the on-demand storage time $T_s$. The Supplementary Methods provide the precise spin-decay parameterization and the write/recall bookkeeping used in the network simulation. In the numerical results below we use $T_{2,\mathrm{AFC}}=307~\mu\mathrm{s}$ and a $10$-ms spin-coherence scale as representative baseline values.

The photon-pair source is described by a two-mode-squeezed-vacuum distribution with mean pair number $\mu$ per temporal mode. For one elementary link of length $L_0$, each telecom photon propagates over $L_0/2$ before reaching the midpoint BSM, giving the single-arm transmission and detection probability
\begin{equation}
\eta_t(L_0)=\eta_{c,t}\eta_{d,t}\,10^{-\alpha L_0/20}.
\label{eq:etat_main}
\end{equation}
The per-mode accepted-BSM probability, denoted $q(\mu,L_0,N)$, includes the source pair-number distribution, memory-write probability, fiber transmission, detector efficiency, and the photon-number-resolving BSM acceptance condition. Its full expression is given in the Supplementary Methods. For $N$ independent temporal modes, the probability that at least one elementary-link event is accepted in one input train is
\begin{equation}
P_0=1-[1-q(\mu,L_0,N)]^N.
\label{eq:P0_main}
\end{equation}

The AFC comb is prepared once and reused for multiple storage trials. If a preparation of duration $T_{\mathrm{prep}}$ remains usable for $T_{\mathrm{storage}}$, the preparation duty factor is
\begin{equation}
D=\frac{T_{\mathrm{storage}}}{T_{\mathrm{prep}}+T_{\mathrm{storage}}}.
\label{eq:D_main}
\end{equation}
One elementary-link attempt consists of the $N$-mode input train, the control sequence, and the round-trip herald latency. We approximate
\begin{equation}
T_{\mathrm{att}}\simeq
NT_{\mathrm{mode}}+T_{\mathrm{ctrl}}+\frac{L_0}{v_f}+T_{\mathrm{proc}},
\label{eq:Tatt_main}
\end{equation}
where $v_f$ is the group velocity in the fiber. The resulting single-link rate is
\begin{equation}
R_0=D\frac{P_0}{T_{\mathrm{attempt}}}.
\label{eq:R0_main}
\end{equation}
Figures~\ref{RATE2}(a) and \ref{RATE2}(b) show $P_0$ and $R_0$ as a function of $L_0$ for $\mu=0.01$ and $0.05$ $f_{\mathrm{clk}}=100$~MHz, with $N$ optimized at each distance.

\subsection{Conditional fidelity of one elementary link}

We evaluate the Bell-state fidelity conditional on an accepted midpoint BSM. The desired event is the sector in which each source generates exactly one pair. For this sector, we describe the energy--time Bell state by the coherence
\begin{equation}
C_{11}=C_{\mathrm{src}}^2V_{\mathrm{HOM}}\lambda_A\lambda_B,
\qquad
F_{11}=\frac{1+\mathrm{Re}\,C_{11}}{2},
\label{eq:F11_main}
\end{equation}
where $C_{\mathrm{src}}$ is the source coherence, $V_{\mathrm{HOM}}$ is the indistinguishability of the two telecom photons, and
$\lambda_A$ and $\lambda_B$ describe any residual phase decoherence of the conditionally retrieved memory qubits. The baseline AFC calculation takes $\lambda_A=\lambda_B=1$; spin-wave dephasing is then treated primarily as a reduction of retrieval efficiency rather than as an independent exponential collapse of the conditional Bell coherence.

Multipair emission and detector noise can also produce accepted BSM events. We therefore write the elementary-link fidelity as
\begin{equation}
F_0=f_{\mathrm{true}}F_{11}+(1-f_{\mathrm{true}})F_{\mathrm{false}},
\qquad F_{\mathrm{false}}=\frac14,
\label{eq:Fmain}
\end{equation}
where $f_{\mathrm{true}}$ is the fraction of accepted events originating from the desired one-pair-per-source sector. The value $F_{\mathrm{false}}=1/4$ is a conservative maximally mixed approximation; the explicit source-sector sum and dark-count terms used to calculate $f_{\mathrm{true}}$ are given in the Supplementary Methods. Figure~\ref{RATE2}(c) shows the resulting single-link fidelity for the two source brightnesses. Increasing $\mu$ increases the link rate, but also increases the relative contribution of multipair sectors and therefore lowers the conditional fidelity.

\subsection{Rate for nested repeater chains}

We next calculate the end-to-end rate for total distance $L_{\mathrm{tot}}$ divided into $n=1$--$4$ equal elementary links, $L_0=L_{\mathrm{tot}}/n$. After one elementary link is established, its memory excitation is retained while the neighboring link continues to generate entanglement. When two neighboring segments are ready, the stored $980$-nm photons at their shared node
are recalled and sent to a local BSM. The corresponding swap success probability
is
\begin{equation}
P_{\mathrm{swap}}(T_L,T_R)
=\eta_{\mathrm{BSM}}^{(980)}\eta_R(T_L)\eta_R(T_R),
\label{eq:Pswap_main}
\end{equation}
where $T_L$ and $T_R$ are the actual storage times of the two recalled excitations and $\eta_R(T)$ contains the age-dependent recall efficiency. Longer buffering times therefore lower the end-to-end rate by reducing the probability of successful recall and swapping. We obtain the nested-chain rates using a renewal Monte Carlo simulation. The simulation generates the elementary links independently, stores successful segments, and attempts a local BSM when two neighboring segments are ready. A failed local BSM consumes only the two participating child segments; already established segments elsewhere in the chain are retained. The
end-to-end entanglement-generation rate is
\begin{equation}
R_{\mathrm{ent}}=\frac{1}{\langle T_{\mathrm{end}}^{(n)}\rangle}.
\label{eq:Rent_main}
\end{equation}
Figure~\ref{RATE2}(d) compares the resulting rates for one to four elementary links. This calculation uses one physical memory per link direction. The analysis above assumes one AFC memory per source direction. At longer distances, this architecture becomes increasingly limited by memory occupancy and the relatively long AFC preparation cycle. Once a link is
successfully heralded, the corresponding memory remains unavailable while waiting for the neighboring link and the subsequent local BSM, preventing the source from being fully utilized.

This limitation can be reduced by connecting each source to a bank of \(K\) independently prepared AFC memories through a \(1\times K\) optical switch. Successive memory photons can then be routed to available memories while previously heralded memories remain stored or other memories undergo re-preparation. Staggering the preparation cycles can further hide much of
the memory-preparation dead time. Such spatial multiplexing therefore reduces source blocking and allows several elementary links to be buffered simultaneously, increasing the probability that compatible links are available for entanglement swapping.

The memory bank can also reduce the average storage time before a local BSM. Because the recall efficiency \(\eta_R(T_s)\) decreases with spin-storage time, younger stored excitations have a larger probability of successful retrieval. Shorter storage times can also improve the conditional fidelity when residual spin dephasing or readout noise is relevant. Thus, a
\(K\)-memory bank is expected to improve the long-distance rate and can also improve the fidelity in storage-noise-limited regimes. These gains must be balanced against the insertion loss and complexity of the additional switching network. The present calculations therefore use the single-memory architecture as a conservative baseline, while a complete event-driven analysis of the \(K\)-memory architecture is left for future work.

\section{Conclusion and outlook}
The device architecture proposed in this article offers several distinct advantages over existing proposals for long-distance entanglement distribution. First, the hybrid integration of YSO is much simpler than diamond-based photonic circuits. Such hybrid integration has already been successfully demonstrated, although for a different purpose. Second, our proposed architecture is based on thin-film lithium tantalate (TFLT), which is inherently stable and is likely one of the only material platforms capable of fully taking advantage of this architecture and hybrid integration approach. We have already demonstrated TFLT nanophotonic circuits with excellent material and device stability, along with device performance suitable for applications in both classical and quantum photonics. Third, the device components used in this article are being actively developed both by our group and by many other research groups, which makes the proposed architecture experimentally realistic and timely. Fourth, for long-distance quantum communication, multiplexing is essential. The proposed device architecture is inherently scalable. A single chip can contain tens to hundreds of such unit cells, and each quantum communication station can include multiple cryogenic modules containing hundreds of unit cells. Therefore, kHz- to MHz-rate true entanglement-based quantum communication could be possible. The main challenges arise from device fabrication, particularly mass manufacturing, testing, and packaging, which are still largely missing at this stage. We believe that the proposed structure can help shift the direction of research in this field, especially in the United States, where much of the focus over the past decade has remained centered on diamond photonics.

\bibliography{Reference}

\end{document}